\documentclass[prl,twocolumn,showpacs,superscriptaddress]{revtex4}

\usepackage{graphicx}
\usepackage{amsmath}
\usepackage{amssymb}
\usepackage{epsfig}

\renewcommand{\v}[1]{{\bf #1}}

\def\eqa{\begin{eqnarray}}
\def\eea{\end{eqnarray}}
\def\be{\begin{eqnarray}}
\def\ee{\end{eqnarray}}

\newcommand{\<}{\langle}
\renewcommand{\>}{\rangle}

\newcommand{\ua}{\uparrow}
\newcommand{\da}{\downarrow}
\newcommand{\ra}{\rightarrow}

\newcommand{\al}{\alpha}
\newcommand{\bt}{\beta}
\newcommand{\del}{\delta}
\newcommand{\Del}{\Delta}

\newcommand{\veps}{\varepsilon}
\newcommand{\ga}{\gamma}
\newcommand{\Ga}{\Gamma}

\newcommand{\la}{\lambda}
\newcommand{\La}{\Lambda}

\newcommand{\si}{\sigma}

\usepackage{color}

\begin{document}

\title{Pairing with dominant triplet component and possible weak topological superconductivity in BiS$_2$-based superconductors}

\author{Yang Yang}
\affiliation{National Laboratory of Solid State Microstructures, Nanjing
University, Nanjing, 210093, China}

\author{Wan-Sheng Wang}
\affiliation{National Laboratory of Solid State Microstructures, Nanjing
University, Nanjing, 210093, China}

\author{Yuan-Yuan Xiang}
\affiliation{National Laboratory of Solid State Microstructures, Nanjing
University, Nanjing, 210093, China}

\author{Zheng-Zao Li}
\affiliation{National Laboratory of Solid State Microstructures, Nanjing
University, Nanjing, 210093, China}

\author{Qiang-Hua Wang}
\affiliation{National Laboratory of Solid State Microstructures, Nanjing
University, Nanjing, 210093, China} \email{qhwang@nju.edu.cn}

\begin{abstract}
We show that the newly discovered BiS$_2$-based superconductors may have a dominant triplet pairing
component, in addition to a subdominant singlet component arising from the spin-orbital coupling. The
pairing respects time-reversal symmetry. The dominant triplet gap causes gap sign changes between the spin-split Fermi pockets. Within a pocket, the gap function respects $d^*_{x^2-y^2}$-wave symmetry, where the star indicates joint spin-lattice rotations. Below the Lifshitz filling level the gap is nodelss and the superconducting state is weak topological. Above the Lifshitz points the gap becomes nodal. The dominant triplet pairing is consistent with the experimental upper critical field exceeding the Pauli limit.
\end{abstract}



\maketitle

Triplet pairing is a special kind of Cooper pairing, most favorably occurs when there are enhanced
ferromagnetic-like spin fluctuations. The reason for the enthusiastic search of triplet pairing is twofold.
On one hand, triplet pairing is only found in He-III\cite{helium} and possibly in Sr$_2$RuO$_4$\cite{sr2ruo4}, but is
otherwise rare in condensed matters. There are singlet-triplet mixtures in noncentrosymmetric materials but the case with dominant triplet component is also rare.\cite{sato} Finding a new material with triplet pairing is therefore a challenge. On the other
hand, the interest in the search of triplet superconductors is enhanced in view of the fact that under a suitable condition
a triplet superconductor may be topologically nontrivial, supporting Majorana modes on the edge or in a vortex core.\cite{ivanov}
The Majorana modes satisfy non-Abelian braiding statistics and thus could be used in topological quantum computing.\cite{ivanov,nayak}

Recently, a new layered superconductor Bi$_4$O$_4$S$_3$ with $T_c \sim 4.5$
K was discovered.\cite{Mizuguchi-1} The conduction plane
is BiS$_2$ layer, with weak inter-layer coupling. The layered
structure is similar to that in cuprates \cite{Cu-SC} and
iron-based superconductors.\cite{iron-SC} Subsequently a family of
superconductors containing the same conduction layer were
discovered. These include LnO$_{1-x}$F$_x$BiS$_2$ (Ln = La, Nd, Ce, Pr,Yb),
\cite{Mizuguchi-2,Demura-1,JhaKumar-56,XingLi-57,JhaSingh-1}, Sr$_{1-x}$La$_x$FBiS$_2$\cite{LinNi-121}, and
La$_{1-x}$M$_x$OBiS$_2$ (M = Ti, Zr, Hf, Th)\cite{YaziciHuang-88}. In such systems the
parent compound is a band insulator or semiconductor. Superconductivity
is realized by electron doping, except in Sr$_{1-x}$La$_x$FBiS$_2$. The basic band structure obtained by first principle
calculations\cite{UsuiSuzuki-44,WanDing-55,Yildirim-16} reveals very interesting features. The Fermi
surface is two-dimensional like and has a good extent of nesting. As
a function of doping, the Fermi surface evolves from X-Y pockets encircling the
midpoints on the edges of the Brillouine zone (BZ) to the $\Ga$-M pockets encircling the
center and the corner of the BZ via a Lifshitz transition. Around this point, the density
of states is higher and would make the system susceptible to various forms of instabilities
unpon the electron-electron interactions. Moreover, there are significant spin-orbital couplings (SOC)
due to the heavy Bismuth atoms.

At present there is no consensus on the cause of Cooper pairing in
these materials. Opinions range from Cooper pairing caused by soft
phonon modes\cite{WanDing-55,Yildirim-16}to pairing caused by strong correlation\cite{hujp,Dagotto}.
As single crystal samples are not available at the present
stage, experimental evidences that would distinguish one versus the other
mechanism are rare. One exception is the experimental
upper critical field (for grain samples) which seems to exceed the Pauli limit, suggesting
triplet pairing.\cite{LiYang-54} The other exception is the large ratio $2\Del/T_c\sim 17$ observed recently,\cite{LiYang-54} implying that the underlying pairing mechanism is likely unconventional. We are thus motivated to investigate the effect of electron correlations. We use the singular mode functional renormalization group (SMFRG) method, which has been successful in many contexts
even when the strength of the electron-electron interaction is intermediate.\cite{wws1,wws2,xyy1,xyy2,xyy3}
The advantage of FRG \cite{wetterich} is the capability of surveying all
electronic instabilities at the same time.\cite{wws1,wws2,xyy1,xyy2,xyy3,frg} The SMFRG developed for fully antisymmetrized interactions
has been recently utilized in the search of correlation driven topological superconductors,\cite{xyy1} and will be applied here given
the SOC.

The main results in this Letter are as follows. We find the pairing in BiS$_2$-based superconductors has a dominant triplet
component, apart from a subdominant singlet component due to SOC. The
pairing respects time-reversal symmetry, and the gap changes sign on spin-split Fermi pockets.
With respect to joint spin-lattice operations (denoted by a superscript $^*$ henceforth), the pairing gap has $d^*_{x^2-y^2}$-wave symmetry, and changes from being nodeless to nodal as the Fermi surface evolves from X-Y pockets to $\Ga$-M pockets around the Lifshitz points. The nodeless $d^*_{x^2-y^2}$-wave phase can be categorized as a time-reversal-invariant weak topological superconducting phase.
The dominant triplet pairing is consistent with the experimental upper critical field exceeding the Pauli limit.

\begin{figure}
\includegraphics[width=8.5cm]{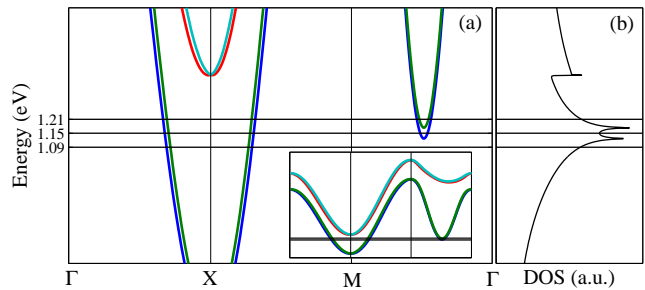}
\caption{(Color online) (a) Spin-split band structure along high symmetry
cuts. The inset is a global view. (b) Normal state density of states. The three horizontal lines
indicate three typical cases to be studied in details. The cusp-like peaks follow from the enhanced density of states at the
Lifshitz points.}\label{band}
\end{figure}

We start by describing the model. Since the inter-layer coupling is weak we consider a one-layer model for brevity, although a unit
cell contains two layers. (We shall come to the effect of inter-layer coupling before closing.)
The conduction bands are mainly derived from the bismuth $p_x$ and $p_y$ orbtals. We write the non-interacting part of the hamiltonian as, $H_0=\sum_{\v k}\psi_{\v k}^\dag (\chi_\v k+\xi_\v k)\psi_\v k$ where $\psi=(C_{x\ua},C_{y\ua},C_{x\da},C_{y\da})^T$
is a four-spinor, with $C_{x/y}$ annihilating the bismuth $p_{x/y}$ orbital. The spin-independent part $\chi_\v k$ can be obtained from Ref.\cite{Mizuguchi-1}, where we observe that the leading second-neighbor hoppings dominates the first-neighbor ones. This feature turns out to have profound consequences in the pairing function (see below). The SOC part can be written as, up to the first neighbors, $\xi_\v k=-\la\tau_2\si_z -\ga_s(\sin k_x\si_y-\sin k_y\si_x)-\ga_d(\sin k_x\si_y+\sin k_y\si_x)\tau_3$. Here $\tau$ and $\si$ are Pauli matrices acting on orbital and spin bases. The $\la$-term is the atomic SOC and the $\ga$-terms are simplest symmetry-allowed Rashba-type couplings (in the absence of mirror symmetry about $z$). A fit to a relativistic band structure calculation~\cite{WanDing-55} suggests $(\la,\ga_s,\ga_d)\sim (0.5,0.02,0.16)$eV. (There is some uncertainty in $\ga_{s,d}$, but it is not crucial in the later development since we get closely similar results by decimating or doubling $\ga_{s,d}$.) The band structure is shown in Fig.\ref{band}(a), where we have four spin-split bands, and the horizontal lines denote the Fermi levels to be considered later. The normal state density of states is shown in Fig.\ref{band}(b), which is enhanced around the Lifshitz points indicated by the arrows.

The interacting part of the hamiltonian we consider is, written in real space, $H_I=U\sum_{i,a}n_{ia\ua}n_{ia\da} +
U'\sum_{i,a>b}n_{ia}n_{ib} + J\sum_{i,a>b,\si,\si'}\psi_{ia\si}^\dag\psi_{ib\si}\psi_{ib\si'}^\dag
\psi_{ia\si'}  + J'\sum_{i,a,b}\psi_{ia\ua}^\dag\psi_{ia\da}^\dag \psi_{ib\da}\psi_{ib\ua}+V\sum_{\<ij\>}n_i n_j.$ Here $a$ and $b$ label the orbitals, $(U,U',J,J')$ are local interactions, and $V$ is the Coulomb interaction between nearest neighbors. We use the standard relation $U=U'+2J$ and $J=J'$ to reduce the number of independent parameters. In the metallic state we expect $V\ll U$, and for the $p$-orbital system we expect $J\ll U$. For simplicity we set $J=V=U/6$, as fine tuning around this setting leads to closely similar results. We have to estimate $U$. Consider partial screening from particle-hole excitations out of the energy window $W$ of the band structure under concern.\cite{crpa} The dielectric constant in this picture is roughly $\veps\sim 1+a V_c/W$, where $a$ is a number of order unity, and $V_c$ is the local unscreened Coulomb interaction for the $p$-orbitals. Assuming $W=5\sim 6$eV, $a=0.5\sim 1$ and $V_c=20\sim 30$eV, we find $V_c/\veps$ ranges from $3$eV to $8$eV. In the absence of a better estimate, we simply take $U\in [2,10]$eV. We stress that the Mott physics is not relevant here for two reasons. First the relevant electron filling is far from half filling per atom per orbital. Second, the $p$-orbitals are fully hybridized so that the interaction projected on the band basis is reduced by roughly a factor of two. This reasoning justifies the use of FRG below.

An interacting system would exhibit collective fluctuations
in particle-particle (pp) and particle-hole (ph) channels, the exact nature of which depends on
the energy scale when they are observed. In our
SMFRG scheme, the fully antisymmetrized effective interaction vertex at a given scale is decoupled
as, \eqa V^{\al\bt;\ga\del}(\v k,\v k',\v q)\ra \sum_m
S_m (\v q) \phi_m^{\al\bt}(\v k,\v q)[\phi^{\ga\del}_m(\v k',\v q)]^*, \eea
either in the pp or ph channels. Here $\al,\bt,\ga,\del$ are
(spin,orbital) labels, $\v q$ is the collective wavevector, and $\v k$ (or $\v k'$) the internal momentum of the
Fermion bilinears $\psi^\dag_{\v k+\v q,\al}\psi^\dag_{-\v k,\bt}$ and
$\psi^\dag_{\v k+\v q,\al}\psi_{\v k,\bt}$ in the pp and ph
channels, respectively. The SMFRG provides coupled flow of all channels versus
a decreasing energy scale $\La$ (the infrared limit of the Matsubara frequency in our case). In the following
we define, in a specific channel, $S(\v q)$ as the leading eigenvalue for a given $\v q$, and $S$ the globally leading one
for all $\v q$. The most attractive and fast growing eigenvalue
represents the dominant ordering tendency with an associated collective momentum $\v Q$ and
eigenfunction (or form factor) $\phi(\v k,\v Q)$, in the respective channel.\cite{note}
The technical details can be found elsewhere.\cite{xyy1}

\begin{figure}
\includegraphics[width=8.5cm]{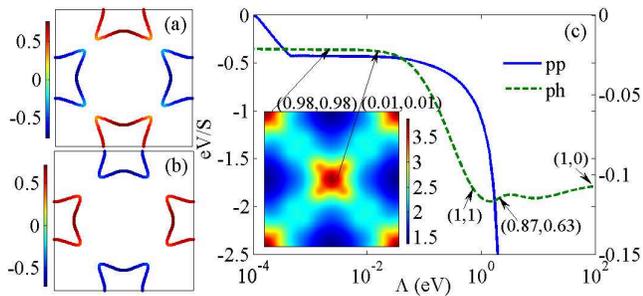}
\caption{(Color online) Results for $n=0.45$. Here (a) and (b) shows the spin-split Fermi surfaces.
The color scale denotes the pairing gap thereon. (c) FRG
flow of $1/S_{\rm pp,ph}$ versus $\La$ associated with the left/right scale. Arrows indicate snapshots of the leading $\v q/\pi$ for $S_{\rm ph}$ during the flow. The inset shows $\ln |S_{\rm ph}(\v q)|$ in the momentum space at the final energy sale.}\label{caseA}
\end{figure}

\begin{figure}
\includegraphics[width=8.5cm]{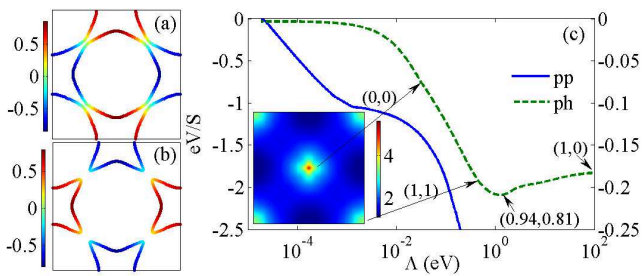}
\caption{(Color online) The same plot as Fig.\ref{caseA} except that $n=0.55$ and $U=4.5$eV.}
\label{caseB}
\end{figure}

First we consider an electron filling $n=0.45$ (per site), a filling level below the Lifshitz points. The Fermi level corresponds to the lowest horizontal line in Fig.\ref{band}. The spin-split Fermi
surfaces are shown in Fig.\ref{caseA}(a) and (b), with electron pockets around X and Y points of the BZ.
For $U=8$eV and $J=V=U/6$, the FRG flow is shown in Fig.\ref{caseA}(c). The interaction
$S_{\rm ph}$ is initially screened slightly, re-enhanced in the intermediate stage and levels off at low energy scales because of
lack of phase space for low energy p-h excitations. The associated momentum evolves (as indicated by the arrows) but eventually settles down on $\v Q\sim(\pi,\pi)$. Inspection of the form factor $\phi_{\rm ph}$ shows that it describes predominantly site-local spin fluctuations. However, from $S_{\rm ph}(\v q)$ (obtained at the final stage of the RG flow) shown in the inset, we see a comparable peak at $\v q=0$, and we find it is also spin-like.
Attractive pairing interaction $S_{\rm pp}$ is induced as $S_{\rm ph}$ is enhanced via the mutual overlap between the channels. There is a cusp in the evolution of $S_{\rm pp}$, which is in fact a level crossing of the leading pairing function $\phi_{\rm pp}(\v k)$. Eventually $S_{\rm pp}$ diverges, implying a superconducting phase. Henceforth we write the matrix pairing function in the orbital basis as $\phi_{\rm pp}(\v k)=(g_\v k+\ga_\v k)i\si_y$, with singlet and triplet parts $g_\v k$ and $\ga_\v k$, respectively. These components can be expressed in terms of lattice harmonics $c_{x/y}=\cos k_{x/y}$ and $s_{x/y}=\sin k_{x/y}$ as well as the Pauli matrices $\tau$ and $\si$. In the present case, we find $g_\v k\sim -(0.03+0.02 c_x c_y)\tau_3$ and $\ga_\v k\sim -0.02(\si_x s_y+\si_y s_x)+0.15(\si_x s_y c_x+\si_y s_x c_y)-0.63\tau_1(\si_x s_x c_y +\si_y s_y c_x)+0.27\tau_3(\si_x s_y c_x-\si_y s_x c_y)$. There are some remarkable features here. First, the singlet part $g_\v k$ transforms as $d$-wave (via $\tau_3$), but the amplitude is relatively small. The dominant part is the $\tau_{1,3}$-terms in $\ga_\v k$ describing triplet pairing on second-neighbor bonds. The reason can be traced back to the dominant second-neighbor hoppings in $\chi_\v k$ mentioned previously. The triplet component is consistent with the existence of small-$\v q$ features in $S_{\rm ph}(\v q)$ discussed above. It is in fact also in harmony with the large-$\v q$ features of $S_{\rm ph}(\v q)$ since on second-neighbor bonds such spin correlations are also ferromagnetic-like. The $d$-wave-like singlet component is also consistent with the large-$\v q$ spin correlations. The singlet and triplet components coexist due to the SOC, and we observe that they transform identically upon {\em joint spin-lattice operations}.
Second, the pairing function respects time-reversal symmetry. This enables us to project $\phi_{\rm pp}$ on the band basis as $\Del_\v k=\<\v k|\phi_{\rm pp}(\v k)(|-\v k\>)^*=\<\v k|g_\v k+\ga_\v k|\v k\>$, where $|\v k\>$ is a Bloch state and $|-\v k\>=T|\v k\>$ is the time-reversal of $|\v k\>$. Thus $\ga_\v k$ would lead to sign changes of $\Del_\v k$ on spin-split bands in the same way as a Rashba-term splits the bands. The subdominant singlet component $g_\v k$ does not alter this conclusion. These observations are exactly reflected in Fig.\ref{caseA}(a) and (b) where $\Del_\v k$ is shown by the color scale. The $d$-wave like sign structure is what we anticipated from inspection of the pairing function in the orbital basis. We dub this kind of pairing symmetry as $d^*_{x^2-y^2}$-wave with the star emphasizing joint spin-lattice operations. We observe that $\Del_\v k$ is nodeless, and the sign changes twice from pockets to pockets (encircling T-invariant momenta). This is in fact classified as a T-invariant weak topological superconducting phase.\cite{qixl} There are gapless edge states in this phase, although they are not topologically protected because of the weak topology. Thus BiS$_2$ may be a good candidate to explore T-invariant topological superconductivity.

We now tune the fermi level to lie in between the Lifshitz points, as indicated by the middle horizontal line in Fig.\ref{band}. The filling level here is $n=0.55$ per site. The spin-split Fermi surface shown in Fig.\ref{caseB} form two hole-like $\Ga$-M pockets (a) and two electron-like X-Y pockets (b). At this filling the density of states is higher, so we use $U=4.5$eV and $J=V=U/6$. ($U>5$ leads to ferromagnetic instability otherwise). The FRG flow is shown in Fig.\ref{caseB}(c). In the ph-channel the momentum settles down at $\v Q\sim 0$ with predominant site-local spin fluctuations. The final $S_{\rm ph}(\v q)$ in the inset is similar to that in the previous case, except the peak at the zone center is much stronger. The existence of such fluctuations would again favor triplet pairing. Indeed, at the divergence scale of $S_{\rm pp}$, we check the pairing function $\phi_{\rm pp}(\v k)$ to find $g_\v k\sim -0.01 \tau _3-0.01i\tau_2 s_x s_y$ and $\ga_\v k\sim 0.11(\si_x s_y c_x+\si_y s_x c_y)-0.58\tau_1(\si_x s_x c_y +\si_y s_y c_x)+0.39\tau_3(\si_x s_y c_x-\si_y s_x c_y)$. The triplet pairing on second neighbor bonds again dominates. Furthermore, the gap projected on the Fermi pockets in (a) and (b) shows that it is again a $d^*_{x^2-y^2}$-wave pairing. However, the gap is now nodal in (a) since the $\Ga$-M pockets are cut by the nodal direction of a $d_{x^2-y^2}$ function. Instead, the gap in (b) remains nodeless on the X-Y pockets which avoid the nodal line.

\begin{figure}
\includegraphics[width=8.5cm]{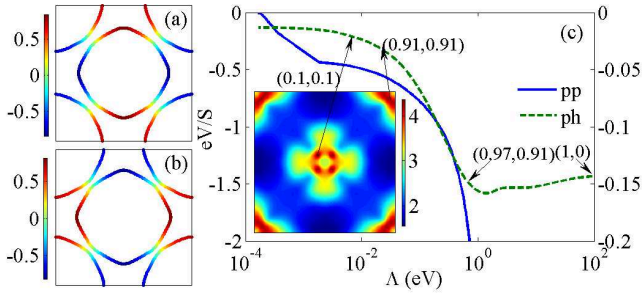}
\caption{(Color online) The same plot as Fig.\ref{caseA} except that $n=0.64$ and $U=6$eV.}
\label{caseC}
\end{figure}

To have better systematics, we further tune the fermi level to lie above the Lifshitz points, as shown by the upper horizontal line in Fig.\ref{band}. The filling level is $n=0.64$ here. The spin-split Fermi surface in Fig.\ref{caseC}(a) and (b) consists of two pairs of $\Ga$-M pockets. For $U=6$eV and $J=V=U/6$, the FRG flow is shown in Fig.\ref{caseC}(c). At the final stage, $S_{\rm ph}$ is associated with a small incommensurate momentum $\v Q\sim (0.1,0.1)\pi$, and we check that it describes predominantly spin fluctuations. The final $S_{\rm ph}(\v q)$ in the inset shows that there are also slightly weaker peaks near the zone corners. The pairing interaction diverges via a final level crossing. The components of the pairing function can be written as, $g_\v k\sim -(0.02+0.06 c_x c_y)\tau_3$ and $\ga_\v k\sim -0.01(\si_x s_y+\si_y s_x)+0.11(\si_x s_y c_x+\si_y s_x c_y)-0.60\tau_1(\si_x s_x c_y +\si_y s_y c_x)+0.36\tau_3(\si_x s_y c_x-\si_y s_x c_y)$. Similar to the previous cases, the triplet component dominates, and the pairing function has $d^*_{x^2-y^2}$-wave symmetry. This is more clearly seen in Fig.\ref{caseC}(a) and (b) for the gap projected on the Fermi surfaces. We notice that because of the comparable large-$\v q$ spin fluctuations, the second leading pairing function has $s_\pm^*$-wave symmetry, with sign changes from $\Ga$- to M-pockets but still with dominant triplet component.

\begin{figure}
\includegraphics[width=8.5cm]{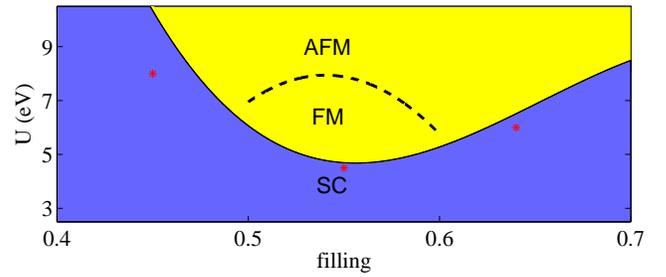}
\caption{(Color online) A schematic phase diagram versus $U$ and electrong filling, assuming $J=V=U/6$. The stars show the cases discussed in the text. Fine tuning around $J=V=U/6$ does not lead to qualitative changes.}\label{pd}
\end{figure}

We have performed systematic calculations for other values of bare interaction parameters and other filling levels. The results are summarized as a schematic phase diagram in Fig.\ref{pd}. The same type of pairing functions are obtained at lower values of $U$, but with a rapidly decreasing divergence scale. Around the Lifshitz points, the system develops ferromagnetism (antiferromagnetism) for $U=5\sim 7.5$eV ($U>7.5$eV). Antiferromagnetism also appears at doping levels away from the Lifshitz points, but only at still larger values of $U$. That the antiferromagnetic state is favored at large $U$ is reflected in the cases examined above: spin fluctuations at large wavevectors are always more efficiently enhanced in the early stage of RG flow. For lower values of $U$, there is a large parameter space favoring the triplet-dominant pairing as discussed above.

Finally we discuss how the results survive in a two-layer model. There will be an inter-layer hopping along the bonds $(\pm 0.5a,\pm 0.5a,\pm b)$ where $a$ (or $b$) is the in-plane lattice constant (or interlayer distance), with  hopping integral $t_\perp$. On the other hand, the Rashba-term reverses sign on the two layers by symmetry. We included these two ingredients and performed SMFRG to find that there are no essential difference in the results. The inter-plane pairing is negligible for a reasonably small $t_\perp$, so we only have to compare the in-plane pairings. We find the only effect in the double-layer model is that the singlet component $g_\v k$ changes sign from one to the other layer.

To conclude, we find that the newly discovered BiS$_2$-based superconductors may have a dominant triplet
component. The pairing respects time-reversal symmetry, and the gap changes sign on spin-split Fermi pockets.
In view of combined spin-lattice rotations, the gap function has a $d^*_{x^2-y^2}$-wave symmetry. It changes from being nodeless to nodal
as the Fermi surface evolves from X-Y pockets to $\Ga$-M pockets around the Lifshitz points. The nodeless $d^*_{x^2-y^2}$-phase is a time-reversal-invariant weak topological superconductor.
The dominant triplet pairing is consistent with the experimental upper critical field exceeding the Pauli limit.

\acknowledgments{QHW thanks Fa Wang, Dung-Hai Lee and Xian-Gang Wan for helpful discussions. The project was supported by NSFC (under grant
No.10974086 and No.11023002) and the Ministry of Science and
Technology of China (under grant No.2011CBA00108 and
2011CB922101).}


\begin{references}
\bibitem{note} Notice that all $\v Q$-vectors have symmetry images, and could evolve with the running energy scale.
\end{references}
\end{document}